**ARTICLE TYPE**

# Near-Infrared Photometry of Superthin Edge-on Galaxies

D. Bizyaev*[1,2,3] | A.M.Tatarnikov[2] | N.I. Shatsky[2] | A.E. Nadjip[2] | M.A. Burlak[2] | O.V. Vozyakova[2]

[1]Apache Point Observatory and New Mexico State University, Sunspot, NM, 88349, USA
[2]Sternberg Astronomical Institute, Moscow State University, Moscow, 119992, Russia
[3]Special Astrophysical Observatory of the Russian AS, 369167, Nizhnij Arkhyz, Russia

**Correspondence**
*Dmitry Bizyaev. Email: dmbiz@apo.nmsu.edu

**Present Address**
This is sample for present address text this is sample for present address text

We perform near-infrared photometry of a large sample of 49 superthin edge-on galaxies. These galaxies are selected based on optical photometry because of high radial-to-vertical scale ratio in their stellar disks. The Near Infrared (NIR) H and K observations were conducted with the cryogenic-cooled camera ASTRONIR-CAM on the 2.5m telescope at the Caucasus Mountain Observatory of Lomonosov Moscow State University. A majority of galaxies in our sample show comparable or better photometric depth than the Sloan Digital Sky Survey (SDSS) optical images. We estimate the structural parameters of stellar disks in the galaxies and find that the NIR scale height of stellar disks is comparable to that estimated from the optical, SDSS $g$, $r$ and $i$, whereas the H and K scale length of the stellar disks is significantly shorter than in the $g$, $r$ and $i$. We investigate if a realistic distribution of dust alone can explain the difference in the scale length and find that in the majority of the galaxies the radial variation of the stellar population is actually responsible for the color distribution. The latter suggests a younger age of the disks periphery, and the inside out building up of stellar disks in the superthin galaxies.

## 1 | INTRODUCTION

The vertical structure of galactic stellar disks can be studied conveniently in galaxies seen at high inclination angles. The dust extinction in the galaxies complicates the studies of the structure. The near-infrared observations help estimate structural parameters of stellar disks less affected by the dust. The galaxies with very small extension in the vertical direction are of special interest. The intensive galactic interaction and merging as well as internal instabilities efficiently heat up galactic disks in the vertical direction (Bournaud et al., 2007; Bullock et al., 2008; D'Onghia et al., 2006; Kazantzidis et al., 2008,0; Khoperskov et al., 2010; Kormendy, 1983; Kormendy & Fisher, 2005; Naab & Burkert, 2003; Purcell et al., 2009; Yoachim & Dalcanton, 2008). These processes are ubiquitous in early evolution of galaxies, and the actual fraction of low redshift galaxies with very thin disks is rather small (Bizyaev et al., 2014,1; Karachentsev et al., 1999; Kautsch et al., 2006; Makarov et al., 2001; Mitronova et al., 2005). They show examples of simplified case of quiet, relatively undisturbed evolution and the disk structure formation (Kautsch, 2009).

Only a few superthin (STG) galaxies (Goad & Roberts, 1981) have been explored in detail up to date. A relatively nearby, prototype superthin galaxy UGC 7321 has been well studied (Matthews et al., 1999; Matthews, 2000; Matthews & Wood, 2001; Uson & Matthews, 2003). Some more large galaxies with superthin disks have been studied individually using kinematic data (Banerjee & Bapat, 2017; Kurapati et al., 2018; Matthews & Uson, 2008; Mendelovitz et al., 2000; O'Brien et al., 2010; van der Kruit et al., 2001) . A few more galaxies with close to the superthin disc scale ratio have been studied in the middle infrared in the frames of the S$^4$G survey (Salo et al., 2015).

The largest sample of superthin galaxies up to date was selected and observed by Bizyaev et al. (2017) (B17 hereafter).

Those galaxies were selected from the Sloan Digital Sky Survey (SDSS), and their structural parameters, such as the stellar disk radial and vertical scales, were estimated from optical images. As it has been noticed before, the structural parameters are affected by the presence of dust, especially in the case of edge-on galaxies (de Grijs et al., 1997; Wainscoat et al., 1989). Although the superthin galaxies are suspected to be under-evolved objects which have less dust than in regular disks (Bizyaev et al., 2017; Kautsch, 2009), the dust extinction cannot be neglected there for proper photometric and spectroscopic modeling. In order to perform modeling of dynamical state of the superthin galactic disks, we need to know the radial and vertical scales of the stellar disks and parameters of the bulges less affected by the dust extinction.

The near-infrared (NIR) photometry helps mitigate the dust effects in estimating the structural parameters of the galactic disks and bulges. This is especially valuable in the case of edge-on galaxies. The superthin galaxies mostly have low surface brightness disks (B17) and bluer stellar population, which makes their NIR observations challenging. Thus, public available 2MASS survey photometry does not have useful data for our superthin galaxies because of insufficient depth.

We used the ASTRONIRCAM camera (Nadjip et al., 2017) on the 2.5-meter telescope at the Caucasus Mountain Observatory of Lomonosov Moscow State University to observe a large set of the superthin edge-on galaxies selected from optical photometric data.

## 2 | OBSERVATIONS AND DATA REDUCTION

Our sample of the superthin galaxies was formed from the same objects considered by B17 and based off of the $g$, $r$ and $i$ SDSS photometry (Gunn et al., 1998; York et al., 2000) and the catalog of edge-on galaxies made by Bizyaev et al. (2014). It includes edge-on galaxies with the radial-to-vertical disk scale ratio greater than 9, which agrees with the original definition of STGs by Goad & Roberts (1981).

The observations were performed on the 2.5m telescope at the Caucasus Mountain Observatory of Lomonosov Moscow State University, using ASTRONIRCAM. The ASTRONIRCAM NIR imager and spectrograph is described by Nadjip et al. (2017) in detail. It is equipped with standard MKO NIR filter set (Legget et al., 2006). The 4.6 arcmin field of view of the camera allows us to dither the NIR frames between the exposures. We used the MKO H and K filters to observe objects from our sample. Each observation was performed as a sequence of short exposures, typically 20-40 seconds each. This provided a balance between the sky background brightness sufficient to be used for simultaneous sky flats, and not saturated centers of the galaxies and bright stars near galactic images.

The large field of view of the camera enabled us to apply spatial dithering between the exposures. The spatial displacement between the dithers was 8-20 arcsec, mostly in the direction along the minor axis. This dithering strategy provides us with the sky flats and the object images simultaneously. The short images were combined to make the H and K sky flats individually for each object. The total integration time was tailored in the way to be able to probe the optical limits of the stellar galactic disks. Typically, we obtained from 40 to 100 short images for each galaxy. Flat fielding and image co-adding was performed in the IRAF standard packages.

The photometric calibration was performed via the H and K magnitudes of non-saturated 2MASS stars identified in the co-added frames. The 2MASS-MKO transformation from Legget et al. (2006) was applied. The wide NIR frames provided us with sufficient number of bright stars in the fields. Typically, we used 7 stars for the cross-calibration against the 2MASS. Our observations are summarized in Table 1. The whole sample includes 49 individual galaxies. A small fraction of NIR images did not get enough photons to show the stellar disks clearly even with long exposures that we performed. This is not surprising because many of superthin galaxies have very blue disks (B17) that emit less than regular galaxies in the NIR.

## 3 | STRUCTURAL PARAMETERS

The images of the galaxies were manually pre-processed in order to determine the centers of the galaxies, their position angle in the NIR frames, and the spatial limits for the model fitting. We estimate the stellar disk scale length, scale height, central surface brightness, and the bulge-to-total ratio using the pipeline that we have run for the SDSS optical images (Bizyaev et al., 2014). The resulting structural parameters are shown in Table 2.

Figure 1 compares the spatial limits of the fitting: the semimajor and semiminor axes of ellipses that encompass each galaxy, determined for each galaxy in the optical (SDSS g, r, and i) and NIR (H and K) images. Since our exposure time varied from object to object, the faintest isophotes in the images look deeper or shallower with respect to their SDSS images. We identify a subsample of 15 our objects whose spatial extension is better than in the optical SDSS images in both directions along the minor and major axes, and whose images are not contaminated by bright stars or artifacts. We call it the "deep" subsample hereafter and use it to verify if insufficient photometric depth can be responsible for the effects that we report. Note that the difference in the semimajor axis of the limiting ellipses between the H and K images is not unexpected since the H images were often made purposely deeper.

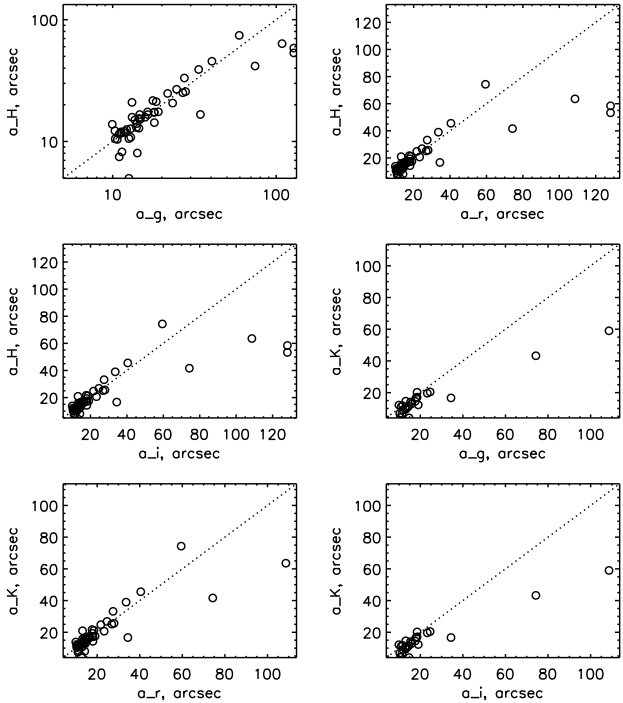

**FIGURE 1** Comparison of the major axis of the limiting ellipse in the NIR (H and K) and optical (g, r, and i) bands.

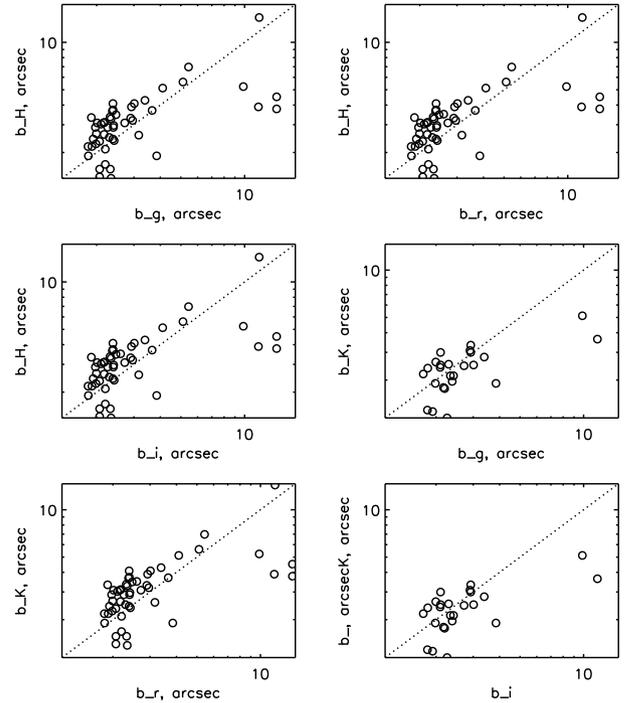

**FIGURE 2** Comparison of the minor axis of the limiting ellipse in the NIR (H and K) and optical (g, r, and i) bands.

Figure 2 shows similar to Figure 1 plots for the comparison of the fitting area size along the minor axis. Figures 1 and 2 demonstrate that the fitting regions of the galaxies in the NIR images are same or more extended along either major or minor axes than for the optical SDSS images for most of observed objects.

The structural parameters of the galactic disks are estimated with the pipeline that was used by Bizyaev et al. (2014). The stellar disk photometric profiles along the major and minor axes were fitted by model profiles taking into account the projection effects. The dust extinction was not modeled, but mitigated by not using the very central pixels along the galactic midplane. The stellar disks were assumed to have exponential luminosity density distribution in the radial "$r$" direction, and isothermal in the vertical "$z$" : $\rho(r,z) = \rho_0(r.z) \, exp(-r/h) \, sech^2(z/z_0)$ (Bahcall, 1984). Here the $h$ and $z_0$ designate the scale length and scale height, respectively. The central luminosity density $I_0 = \int \rho(0,z)dz$ of the face-on disk is converted to the surface brightness $\mu_0$ via the flux calibration procedure described in §2. The structural parameters of the galactic disks are summarized in Table 2.

Although alternative functional forms of the vertical distribution of light and matter in disks are traditionally considered (Banerjee & Jog, 2007; Kurapati et al., 2018; Pohlen et al., 2000; van der Kruit, 1988), there are some reasons that keep us using the same "sech$^2$" function for all galaxies in our sample for this study. First, enough pixels in the vertical direction are required to distinguish between different functions, which is available only in a few largest galaxies from our NIR superthin sample. More over, using the same function for the NIR data as has been used for the optical sample makes the comparison of the vertical scales straightforward.

Figure 3 and Figure 4 compare the radial and vertical scales of stellar disks in the optical (SDSS) and our NIR images. It can be seen that the vertical scale height rather similar between the g,r,i and H,K images. The NIR vertical scale heights are typically 0.9 of the optical scales determined form the g,r,i images, in agreement with Bizyaev & Mitronova (2009); Zasov et al. (2002). At the same time, the small difference between the observed optical and NIR vertical scales for the superthin disks points at their lower extinction than in general edge-on galaxies Bizyaev & Mitronova (2009). In the latter case we expect to see a bigger difference between the optical and NIR vertical scales.

The estimated radial scale length in the NIR is approximately twice as less as that in the optics. We obtain similar results with the general and deep NIR subsamples. The scale ratios between different bandpasses are summarized in Table 3, where we show the mean, median, and standard deviation values of the ratios of the structural parameters.

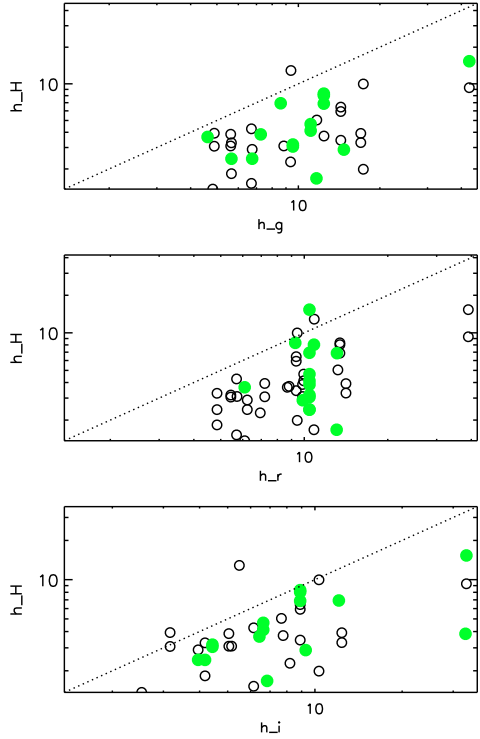

**FIGURE 3** Comparison of the stellar disk scale length in the H and optical bands. The green bullets designate the deep subsample of galaxies.

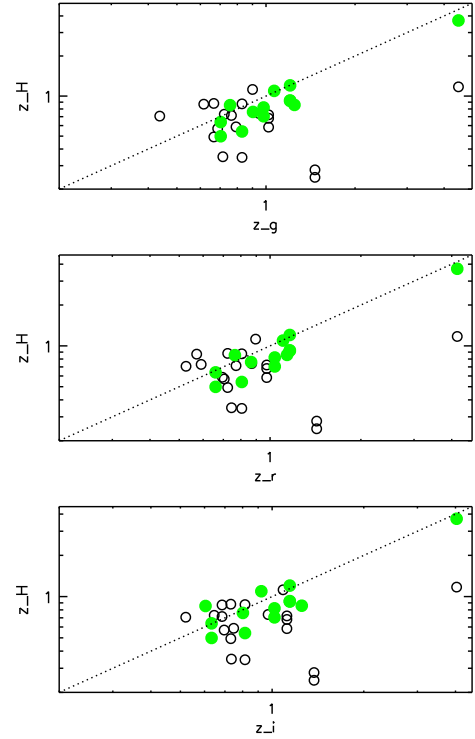

**FIGURE 4** Comparison of the stellar disk scale height in the H and optical bands. The green bullets designate the deep subsample of galaxies..

The main conclusion from the comparison between the NIR and optical images is that the vertical scale heights keeps almost the same value between the NIR and optics, while the radial scales are significantly shorter in the NIR, in agreement with de Grijs (1998); Zasov et al. (2002). It was noticed that the superthin galaxies do not make a uniform group (B17): the red and the blue galaxies have different properties and may represent different evolutionary paths. The two groups can be distinguished by their optical color , e.g. (r-i) (B17). We do not see any trends in difference between the NIR and optical scale lengths in dependence of the galactic color.

Possible explanations for the large difference of the radial scales can come either from the radial variation of the dust extinction with no systematic change in the stellar population, or from true radial gradients of the stellar populations. The former possibility is considered below.

## 3.1 | Can the dust extinction explain the shorter NIR disk scale length?

A special distribution of dust in galactic disks can potentially, be a reason for the difference between the optical and NIR scale length that we observe in our galaxies. We can figure out what kind of the radial dust distribution is required to explain our observing color distributions in the NIR in combination with optical distributions from SDSS. Note that we neglect the dust scattering because it affects the color distributions on a short spatial scale and does not affect the structural parameters of the disks significantly (Bianchi, 2007).

We are able to figure out the radial distribution of $A_V$ by considering the NIR and optical colors simultaneously. We use only g,r, & i SDSS color combination because typically the u and z SDSS images are more noisy and have less depth. For the true $(a-b)_0$ and observed $(a-b)$ colors in two generalized bands $a$ and $b$ one can write $(a-b) = (a-b)_0 + A_a - A_b$. Here we use the available filters g,r,i,K and H as the $a$ and $b$. The color excess ratios $C_a$ make connections between the extinction $A_a$ in the filter $a$ and $A_V$ in the V like $A_a = C_a A_V$. We adopt the NIR and SDSS bands color excess ratios from (Schlafly & Finkbeiner, 2011). The relationship between the extinction and colors can be written

$$(a - b) = (a - b)_0 + (C_a - C_b) A_V \qquad (1)$$

First, we consider a case if our galaxies had no radial variations in the stellar population. This corresponds to the equal scale length in all, optical and NIR, bands and anticipates that the true radial distribution of colors is flat. With this assumption we can figure out the $A_V$ from eq. (1). Figure 5 (bottom panel) shows an example of $A_V$ estimated from the (g-H), (r-H), and (i-H) colors for one galaxy of our deep subsample for $R_V =$ 3.1. The difference between the radial distributions of $A_V$ is typical for most of the galaxies - the $A_V$ from different color combinations do not agree between each other and show different slopes, which suggests that there is no combination of true, non changing colors that can explain the observed colors with common extinction. In 3 galaxies out of 15 we notice the radial trends of the $A_V$ as expected with the same stellar population at different regions in the galactic disks. Since this is a minor fraction, we conclude that having the same stellar population at different regions throughout the galactic disks is a possible, but not a common case, and this cannot explain the difference between the optical and NIR disk scale lengths in general.

Next, we attempt to estimate the $A_V$ from the combination of colors using the equations above. In addition to eq. (1), we notice that the evolutionary sequence of stellar populations for different ages and metallicities is degenerated in the plane of SDSS g,r,i colors. The sequence of simple stellar population models from Bruzual & Charlot (2003) for metallicities between the solar and 10 times metal poor, and the ages between 10 and 0.1 Gyrs forms a narrow, line-like zone on the $(g-r)_0 - (g-i)_0$ diagram. The intrinsic color relationship there can be parameterized as $(g-r)_0 = 0.714(g-i)_0 - 0.043$. Combining this equation with eq. (1), we can solve the system of equations for $A_V$ and true NIR - optical colors in the sense of $(H, K-g, r, i)_0$ in each point along the galactic radius independently.

The top panel in Figure 5 shows the observing radial distribution of colors for one of our objects. The colors were estimated along the major axis of all galaxies, in the same radial bins 1 arcsec wide. The second top panel in Figure 5 demonstrates the $A_V$ estimated for three cases of $R_V$: 2.1 (green dotted curve), 3.1 (solid blue), and 4.1 (dashed red). The $R_V =$ 3.1 (Cardelli et al., 1998) is prefered by Schlafly & Finkbeiner (2011), and we try the other $R_V$ values for illustration purposes. Note that we did not restrict the reddening by positive values only, and at certain points the computed $A_V$ can be negative, which points at problems with the initial assumption in this case. In addition to some negative values, the $A_V$ in some galaxies does not decrease systematically at large radii. From spectral observations of many galaxies we expect to see negative radial gradients of gas phase metallicity as well as a decreasing of the dust extinction at periphery (Belfiore et al., 2017; Sanchez et al., 2014; Sanchez-Menguiano et al., 2016). Having the extinction getting systematically higher with radius is unlikely, unless an exotic radial distribution of gas density, metallicity, or dust-to-gas ratio is anticipated.

The third top panel in Figure 5 displays the true, intrinsic colors $(g-H)_0$, $(r-H)_0$ and $(i-H)_0$ obtained from the equations above. The NIR - optical color distributions show the descending radial trends, as expected, and cannot be called constant. Among the 15 deep galaxies in all color combinations only 4 indicate a non-negative, non-rising radial distributions of $A_V$ in a combination with small, non-systematic observed same deep or deeper than the SDSS images. We estimate the NIR structural parameters: the vertical and radial scales of stellar disks, and compare them to the optical parameters. While the vertical scale heights are comparable in the NIR and optics, the radial scale length shows significant difference: the galactic disks have longer scales in the blue bands, and short scales in the NIR. We investigate if a realistic radial distribution of dust can cause the difference between the optical and NIR scale length. We conclude that the extinction variation can explain the scale length difference for only a small fraction (under 20%) of the galaxies. In all others a radial variation of the integral stellar population is required, which means the observed scale length difference between the NIR and optical bands is real. This anticipates a younger age and low metallicity of the disks periphery. In agreement with spectral studies of regular disk galaxies, it is compatible with the inside out building up of stellar disks in superthin galaxies.

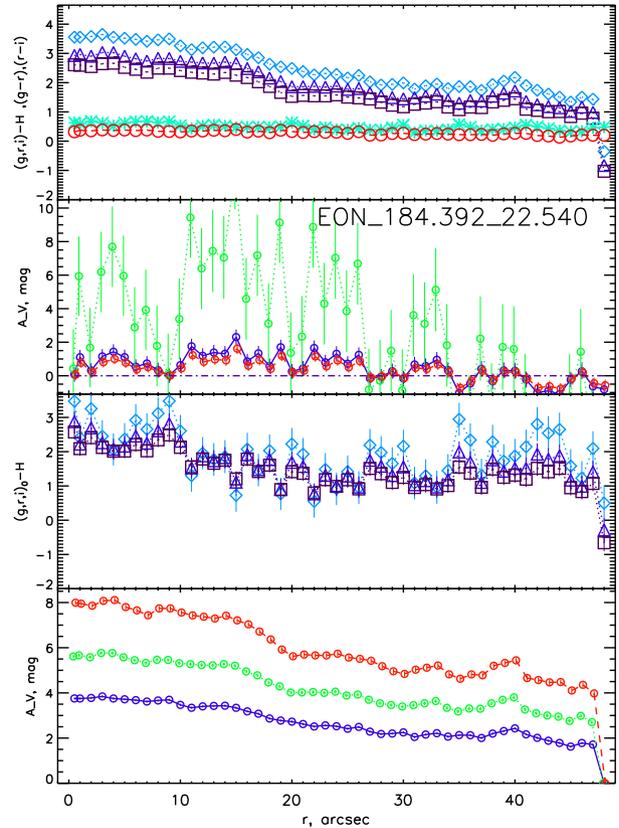

**FIGURE 5** Results of the extinction modeling for one of the galaxies (EON_184.392_22.540), which do not agree with the equality of the NIR and optical scale length. Top: the observed radial distributions of (g-K) - blue diamonds, (r-K) - navy triangles, (i-K) - purple squares, (g-r) - red circles, and (g-i) - teal asterisks. Second top: the calculated radial distributions of the extinction $A_V$ for the case of $R_V = 2.1$ (green dotted curve), 3.1 (blue solid curve), and 4.1 (red dashed curve), see text. Third top: the (g-K), (r-K) and (i-K) colors (blue diamonds, navy triangles and purple squares, respectively) corrected using the extinction above for the case of $R_V = 3.1$. Bottom: the $A_V$ calculated for the case of flat distribution of the colors (g-K), (r-K) and (i-K) designated with the solid blue, green dotted, and red dashed curves, respectively. The error bars in the plots correspond to $1\sigma$ errors.

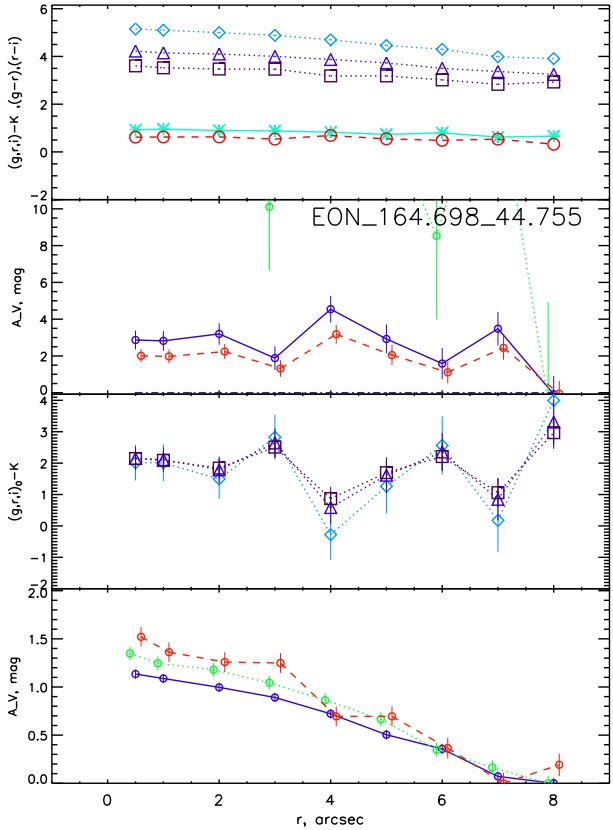

**FIGURE 6** Results of the extinction modeling for one of the galaxies (EON_164.698_44.755), which are consistent with similar scales in the NIR and optical bands. The designations are kept the same as in Figure 5.

(with respect to uncertainties) variations of the intrinsic NIR-optical colors in the second and third top panels in Figure 5. An example of such a galaxy is shown in Figure 6. Since this combination would point at the dust distribution as an acceptable explanation of the difference between the optical and NIR scale lengths, we conclude that this is a possible, but not a typical case for our sample of galaxies. Instead, a real radial variation of integral stellar population contributes to the observing NIR and optical color distributions and causes the difference between the NIR and optical disk scale lengths.

## 4 | CONCLUSIONS

We report the NIR observations of 49 superthin edge-on galaxies selected by Bizyaev et al. (2017) from the optical SDSS photometry. Significant fraction of our sample is

**Table 1.** Observations of edge-on galaxies

| Object | Date | Filter | Exp time | FWHM |
|---|---|---|---|---|
| EON_7.069_24.845 | 03feb2016 | K | 1080 | 2.1 |
| EON_15.871_15.475 | 03feb2016 | K | 1080 | 1.7 |
| EON_17.275_19.605 | 12dec2015 | H | 1943 | 1.4 |
| EON_19.768_-0.139 | 30oct2015 | H | 1862 | 1.6 |
| EON_23.581_53.734 | 09dec2015 | H | 2286 | 1.1 |
| EON_29.178_17.711 | 07mar2016 | H | 1720 | 2.0 |
| ... | | | | |

**Table 2.** The NIR structural parameters of galactic disks.

| Object | Filters | $h$ | $dh$ | $z_0$ | $dz_0$ | $\mu_0$ | $d\mu_0$ |
|---|---|---|---|---|---|---|---|
| | | arcsec | arcsec | arcsec | arcsec | mag/arcsec$^2$ | mag/arcsec$^2$ |
| EON_7.069_24.845 | K | 3.81 | 0.12 | 0.71 | 0.09 | 17.75 | 0.50 |
| EON_15.871_15.475 | K | 5.28 | 1.21 | 0.75 | 0.22 | 20.05 | 0.42 |
| EON_17.275_19.605 | H | 5.21 | 0.25 | 1.05 | 0.16 | 21.25 | 0.26 |
| EON_19.768_-0.139 | H | 14.85 | 1.89 | 2.12 | 0.19 | 20.53 | 0.17 |
| EON_23.581_53.734 | H | 5.27 | 0.83 | 1.01 | 0.04 | 18.86 | 0.35 |
| EON_29.178_17.711 | H | 9.98 | 0.68 | 0.89 | 0.11 | 20.37 | 0.22 |
| ... | | | | | | | |

**Table 3.** Comparison of optical and NIR structural parameters of galactic disks

| Parameter | Filters | Mean | Median | Std. dev. |
|---|---|---|---|---|
| $h$ | H/g | 0.464 | 0.429 | 0.179 |
| $h$ | K/g | 0.516 | 0.565 | 0.201 |
| $h$ | H/r | 0.489 | 0.469 | 0.171 |
| $h$ | K/r | 0.445 | 0.473 | 0.153 |
| $h$ | H/i | 0.626 | 0.658 | 0.230 |
| $h$ | K/i | 0.564 | 0.522 | 0.303 |
| $z_0$ | H/g | 0.856 | 0.827 | 0.289 |
| $z_0$ | K/g | 0.931 | 0.920 | 0.198 |
| $z_0$ | H/r | 0.870 | 0.845 | 0.275 |
| $z_0$ | K/r | 0.925 | 0.962 | 0.187 |
| $z_0$ | H/i | 0.863 | 0.811 | 0.278 |
| $z_0$ | K/i | 0.961 | 1.037 | 0.199 |
| Same ratios for the deep subsample | | | | |
| $h$ | H/g | 0.494 | 0.429 | 0.190 |
| $h$ | K/g | 0.622 | 0.608 | 0.142 |
| $h$ | H/r | 0.481 | 0.500 | 0.099 |
| $h$ | K/r | 0.487 | 0.393 | 0.166 |
| $h$ | H/i | 0.612 | 0.622 | 0.224 |
| $h$ | K/i | 0.549 | 0.560 | 0.278 |
| $z_0$ | H/g | 0.836 | 0.827 | 0.144 |
| $z_0$ | K/g | 0.877 | 0.916 | 0.115 |
| $z_0$ | H/r | 0.855 | 0.797 | 0.138 |
| $z_0$ | K/r | 0.891 | 0.884 | 0.150 |
| $z_0$ | H/i | 0.906 | 0.811 | 0.217 |
| $z_0$ | K/i | 0.964 | 1.037 | 0.182 |

## ACKNOWLEDGMENTS

This research is based on observational data obtained at the Caucasus Mountain Observatory of Lomonosov Moscow State University. This work was supported in part by M.V. Lomonosov Moscow State University Program of Development. D.B. is partly supported by grant RScF 19-12-00145.